\documentclass{article}
\usepackage{spconf,amsmath,graphicx,url,times,bm}
\usepackage{color}

\usepackage{type1cm,bm,amsmath,amssymb}
\usepackage{breqn,multirow,autobreak,booktabs}
\title{Real-time Stereo Speech Enhancement with Spatial-Cue Preservation based on Dual-Path Structure}
\name{Masahito Togami, Jean-Marc Valin, Karim Helwani, Ritwik Giri, Umut Isik, and Michael M. Goodwin}
\address{
Amazon Web Services, Palo Alto, CA, USA}
\begin{document}
\ninept
\maketitle
\begin{abstract}
We introduce a real-time, multichannel speech enhancement algorithm which maintains the spatial cues of stereo recordings including two speech sources. Recognizing that each source has unique spatial information, our method utilizes a dual-path structure, ensuring the spatial cues remain unaffected during enhancement by applying source-specific common-band gain. 
This method also seamlessly integrates pretrained monaural speech enhancement, eliminating the need for retraining on stereo inputs. Source separation from stereo mixtures is achieved via spatial beamforming, with the steering vector for each source being adaptively updated using post-enhancement output signal. This ensures accurate tracking of the spatial information. The final stereo output is derived by merging the spatial images of the enhanced sources, with its efficacy not heavily reliant on the separation performance of the beamforming. The algorithm runs in real-time on 10-ms frames with a 40 ms of look-ahead. Evaluations reveal its effectiveness in enhancing speech and preserving spatial cues in both fully and sparsely overlapped mixtures.

\end{abstract}
\section{Introduction}
\label{sec:intro}

Industrial speech communication systems, e.g., teleconferencing systems, are often used in noisy and reverberant environments, so speech enhancement techniques are needed to ensure clear communication  \cite{loizou2013,se2005}.  In production grade teleconferencing applications, enabling two-way communication imposes constraints of real-time processing and low input/output latency.  Additionally, stereo input is becoming increasingly common in current teleconferencing situations, so speech enhancement methods for two-channel input are needed.  In such solutions, it is important not only to achieve high speech enhancement performance 
but also to preserve the spatial cues of speech sources, because spatial-cue information is 
an important clue to know who is speaking. 

Many approaches for monaural speech enhancement have been studied \cite{loizou2013,mmse1984,ss1979}, e.g. spectral subtraction \cite{ss1979} and  Bayesian methods such as minimum mean-square error (MMSE) methods \cite{mmse1984}.   However, the speech enhancement performance of MMSE methods is not sufficient for nonstationary noise. 
Recently, deep neural network (DNN) based methods have been applied for monaural speech enhancement or speech source separation \cite{Xu2015,Erdogan2015,deep_clustering1,PIT2017,pascual17_interspeech,ConvTasNet2019}. 
Due to the strong expression capability of DNNs for speech characteristics, the performance of monaural speech enhancement algorithms for nonstationary noise conditions has improved dramatically.

 While research on DNN-based speech enhancement has focused on monaural algorithms, multichannel speech enhancement which combines multichannel spatial beamforming and DNN has also been recently explored in an effort to improve on established classical methods  \cite{heymann2016, Higuchi2018}. Recently, methods which incorporate DNNs more directly into multichannel processing have been studied, e.g., \cite{MISD2019,masuyama19_interspeech,Han2020, Bahareh2022}.  However, it is necessary to train a specific DNN model for multichannel signals. 

In parallel with the study of DNN-based multichannel speech enhancement, multichannel speech enhancement techniques with spatial-cue preservation based on 
traditional signal processing frameworks have been also actively studied  \cite{van_den_bogaert2007,szurley2016,andreas2017,klasen2005,cornelis2010,Costa2014,Tha21-CG}. The common-gain method \cite{Tha21-CG} is a straightforward speech enhancement approach with spatial-cue preservation for single-source cases. It estimates a common time-frequency gain for all microphone input signals. The common-gain based method ensures preservation of the interaural phase difference (IPD) and interaural level difference  (ILD) between channels.

In this paper, we propose a real-time stereo speech enhancement method with spatial-cue preservation under the condition that there are two speech sources. 
Considering that spatial information such as phase difference and amplitude ratio  between microphones is different for each source, 
the proposed method separates two speech sources by multichannel spatial beamforming followed by DNN-based monaural speech enhancement with a common gain specific to each source. As the monaural speech enhancement, the proposed method adopts  the state-of-the art PercepNet algorithm \cite{valin2020perceptually} which operates in real-time on 10-ms frames with 30 ms of look-ahead, and there is no need to train a stereo-specific DNN model. 
The steering vector of each source required for spatial beamforming is adaptively updated using the output signal after speech enhancement to track spatial information of the speech source rapidly.  A final stereo signal is obtained by remixing spatial images of the enhanced speech sources.
We evaluate the proposed method using two datasets with different speech overlap characteristics.

\section{Problem statement}
\subsection{Signal model}
In our setup, we consider the scenario of two microphones capturing the sound of two speech sources in a noisy environment.  
The $m$-th microphone input signal $x_{m,t}$  ($t$ is the time-index) is modeled in the time domain as follows:
\begin{equation}
x_{m,t}= \sum_{i=1}^{2} s_{i,t} \ast h_{i, m}+ n_{m,t}, 
\end{equation}
where $s_{i,t}$ is the $i$-th speech source signal, $h_{i,m}$ is the impulse response between the $i$-th speech source location 
and the $m$-th microphone location, and $n_{m,t}$ is the background noise signal. Speech enhancement is carried out in the time-frequency domain.
The time-frequency representation of $x_{m,t}$ can be written as follows:
\begin{equation}
\bm{x}_{l,k}=\sum_{i=1}^{2} s_{i,l,k}\bm{a}_{i,k}+\bm{n}_{l,k},
\end{equation}
where $l$ is the frame-index, $k$ is the frequency index, $\bm{x}_{l,k}=[\begin{array} {cc} x_{1,l,k}  & x_{2,l,k}  \end{array}]^{T}$, $T$ is the transpose operator of a matrix or a vector, and $x_{m,l,k}$ is the time-frequency representation of the time-domain signal $x_{m,t}$.  $\bm{n}_{l,k}$ is defined  as $\bm{n}_{l,k}=[\begin{array} {cc} n_{1,l,k}  & n_{2,l,k}  \end{array}]^{T}$. 
$\bm{a}_{i,k}=[\begin{array}{cc}a_{i,1,k}  & a_{i,2,k} \end{array} ]^T$  is a steering vector and $a_{i,m,k}$ is the time-frequency representation of $h_{i,m}$. 

In this paper, we focus on stereo speech enhancement with spatial-cue preservation. Thus, the objective is defined as extraction of $\sum_{i=1}^{2} s_{i,l,k}\bm{a}_{i,k}$ from the microphone input signal $\bm{x}_{l,k}$. We assume that a pretrained DNN-based monaural speech enhancement method is incorporated in the system. For the experiments in this paper, the state-of-the art PercepNet method \cite{valin2020perceptually} is used for monaural speech enhancement. PercepNet satisfies our 
design requirements in for real-time, low-latency, high-quality enhancement; it operates on 10-ms frames with 30 ms of lookahead, and ranked second in the real-time track of the recent DNS challenge despite operating at much lower than the allowed complexity \cite{dns2020}.
% We utilize the original PercepNet with no modification. 

\subsection{Discrete channel processing}
\begin{figure}[htb]
	\centering
	\includegraphics[width=0.7 \linewidth]{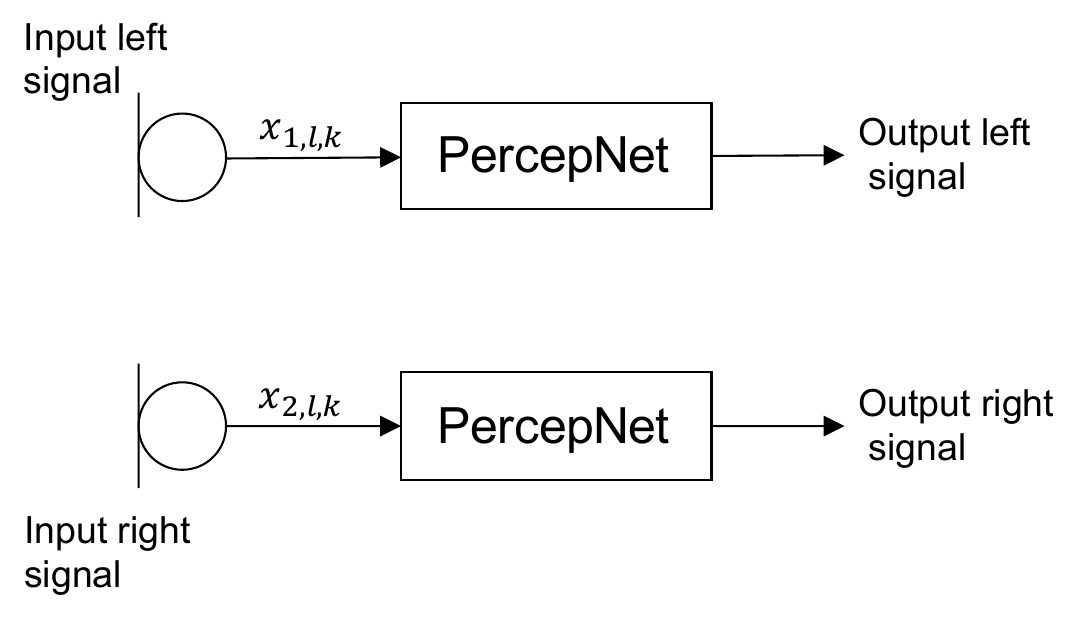}
	\caption{Block diagram of discrete-channel processing method. }
	\label{fig:discrete_processing}
	\vspace{-20pt}
\end{figure}

One approach for enhancing a stereo speech signal is to perform monaural speech enhancement independently for each channel (Fig.~\ref{fig:discrete_processing}).  We use this discrete-channel processing approach as a baseline for assessing our algorithm.   In discrete-channel processing, it is not guaranteed that each speech source is enhanced the same way in each output channel, and as a result the output spatial image can be unstable. 

\subsection{Common gain method} 
\begin{figure}[htb]
	\centering
	\includegraphics[width=0.8 \linewidth]{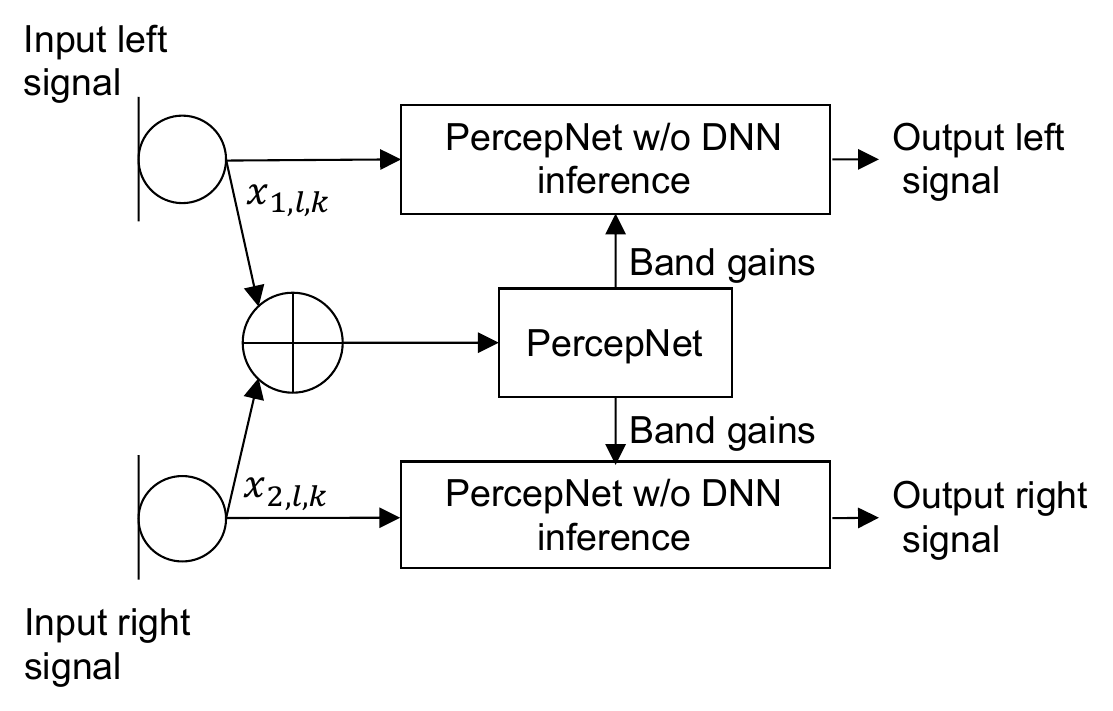}
	\caption{Block diagram of common gain method. }
	\label{fig:common_gain}
	\vspace{-20pt}
\end{figure}

An alternate method is proposed in \cite{Tha21-CG} where a common time-frequency gain for both microphone signals is estimated (Fig.~\ref{fig:common_gain}). The output signal of each channel is obtained by multiplying the common time-frequency gain by the input signal of the corresponding channel. The common-gain method ensures that the ILD and IPD of the output stereo signal are the same
as those of the microphone input stereo signal. PercepNet operates on only triangular spectral bands, spaced according to the equivalent rectangular bandwidth (ERB) scale rather than time-frequency representation, and each band energy of the output signal is estimated via multiplying the estimated band gain by the band energy of the input signal. Thus, the common gain method is applied to the ERB scale in this paper. 
In our implementation of this approach depicted in Fig.~\ref{fig:common_gain}, the common band gain is calculated by PercepNet operating on a downmix of the left-channel input signal and the right-channel input signal.  PercepNet 
estimates $N$ band gains, where $N$ is set to be smaller than the number of frequency bins
in the input time-frequency representation. The band gains derived from the downmix are then shared between channels. In each channel,
it is not needed to run DNN inference to estimate band gains and PercepNet w/o DNN inference is performed.  We use this common gain method as a baseline method in evaluation. 

\section{Proposed framework}
\subsection{Overview}
\begin{figure*}[htb]
	\centering
	\includegraphics[width=0.7 \linewidth]{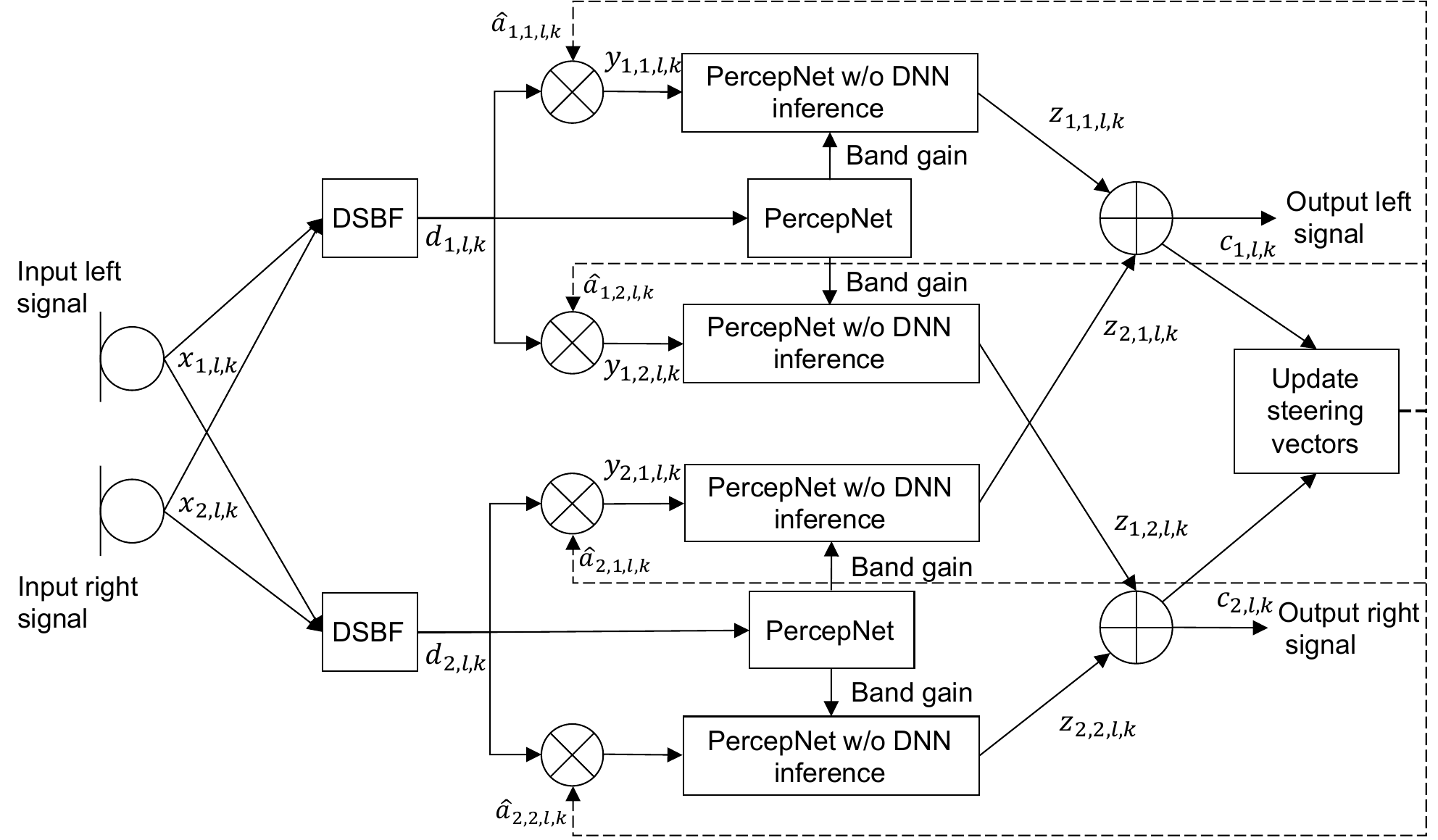}
	\caption{Block diagram of proposed stereo speech enhancement algorithm. }
	\label{fig:proposed_overview}
	\vspace{-10pt}
\end{figure*}

An overview of the proposed framework is shown in Fig.~\ref{fig:proposed_overview}.  
The approach is designed to achieve high-quality speech enhancement using spatial selectivity
as well as monaural speech enhancement.  
To ensure robustness in the presence of two speech sources, we use 
a dual-path structure with two instances of spatial beamforming based on delay-and-sum beamformer (DSBF) and the common-gain method. In the first path 
(the top half), a dominant speech source is enhanced.  In the second path (the bottom half),
another speech source is enhanced. Separating the speech sources initially
via DSBF improves the signal-to-noise ratio of the input signal for the 
monaural speech enhancement and correspondingly improves the quality of the output signal.
The DSBF steering vector is determined from the outputs of the monaural 
speech enhancement model and fed back to the beamformer, as will be explained later.
In each path, the DSBF output signal is used to serve as the input to common-gain estimation.
Similar to the common gain method, it is not needed to run DNN inference to estimate band gains in each channel and PercepNet w/o DNN inference is
performed. A spatial image of the enhanced speech signal 
for each channel is used as the input of the PercepNet w/o DNN inference 
instead of the microphone input signals; this provides 
improved speech enhancement performance. 
After the two paths are processed, the output signal for each channel is generated by remixing 
the respective output channel signals of both paths. 
Note that under the assumption that the monaural speech enhancement algorithm provides a
distortion-free speech output, the dual-path system ensures that two speech sources are perfectly reconstructed. 
The proposed method operates on 10-ms frames, and there is  30 ms of look-ahead in PercepNet and additional 10 ms look-ahead exists in stereo processing outside of PercepNet. Thus, the proposed method operates on total 40 ms of look-ahead. 

% In order to perform high-quality speech enhancement using spatial information as well as monaural 
% speech enhancement, the proposed method uses an enhanced signal by a  delay-and-sum beamforming 
% (DSBF) using a steering vector obtained via DNN as an input signal for the common band gain 
% estimation. In the common gain estimation of the proposed method, a spatial image of the enhanced 
% speech signal for each channel is utilized instead of the microphone input signal of the 
% corresponding channel to improve speech enhancement performance. 
% In order to respond to an environment where multiple speech sources exist, we apply a dual-path 
% structure in which two types of the common gain method is used. In the first path, a dominant source 
% is enhanced and the other sources are enhanced in the second path. By separating the speech sources 
% by spatial beamforming in advance, it can be expected that signal-to-noise ratio of the input signal % for  monaural speech enhancement is improved and the quality of the output signal also improves.
% After that, the output signal of each channel is generated by remixing of the output signals of the 
% corresponding channel of both paths. 
% Importantly, under the assumption that monaural speech enhancement outputs speech signal with no 
% distortion, it is assured that multiple speech sources are perfectly reconstructed in the proposed 
% dual-path structure. 

\subsection{Spatial beamforming with DSBF}
Let $\hat{\bm{a}}_{1,l,k}$, $\hat{\bm{a}}_{2,l,k}$ be the estimated steering vectors of the first path and the second path, respectively. These steering vectors are updated in the previous frame, as described later. 
The input signal for the monaural speech enhancement for each path $d_{i,l,k}$ is obtained by using the DSBF as follows:
\begin{equation}
	d_{i,l,k}=\hat{\bm{a}}_{i,l,k}^{H} \bm{x}_{lk},
\end{equation}
where $H$ denotes the Hermitian transpose operator. 
The $m$-th input signal for the $i$-th path, $y_{i,m,l,k}$, is estimated as a spatial image of the enhanced speech signal of the $i$-th path as follows:
\begin{equation}
y_{i,m,l,k} = d_{i,l,k} \hat{a}_{i,m,l,k}.
\end{equation}

\subsection{Monaural speech enhancement} 
For each path, PercepNet estimates a common band gain from $d_{i,l,k}$. 
In each path, band gains are shared between channels so as to preserve spatial information specific to each source.  The output signal after monaural speech enhancement for the $m$-th channel of the $i$-th path $z_{i,m,l,k}$ is obtained. The final output signal for each channel $c_{m,l,k}$ is obtained as $
c_{m,l,k} = \sum_{i=1}^{2} z_{i,m,l,k}.$

\begin{table*}[htb]
	\caption{Evaluation results}
	\label{tab:eval_result}
	\centering
	\begin{tabular}{l  r  c c c c c c}
		\toprule
		   & \textbf{Dataset}  & \multicolumn{3}{c}{\textbf{WSJ1}} & \multicolumn{3}{c}{\textbf{Stereo Sparse Librimix}} \\ \midrule
		\textbf{Single / Dual} &  &  \textbf{IPD error} &  \textbf{ILD error} & \textbf{MOS} &   \textbf{IPD error} &  \textbf{ILD error} & \textbf{MOS}  \\
		\midrule  
		& Discrete-channel processing & 0.234 &	3.53 & 	3.27 $\pm$ 0.06 & 	0.192 & 	3.68 &	3.47 $\pm$ 0.06 \\ \midrule
	\multirow{3}{*}{Single}	& Common gain (Baseline) & 0.232 & 	3.03 & 	3.27 $\pm$ 0.06 & 	0.194 & 	3.22 & 	3.48 $\pm$ 0.06 \\
		& Common gain (Proposed) & 0.207 & \textbf{1.93} & 	3.29 $\pm$ 0.07 & 	0.158 & 	\textbf{1.63} &	3.48 $\pm$ 0.06 \\ \midrule
	\multirow{3}{*}{Dual}	& Common gain w/ NSV & 0.239 & 	2.64 & 	3.26 $\pm$ 0.06 & 	0.187 & 	2.72 & 	3.51 $\pm$ 0.06\\ 
		& Common gain (Proposed) & \textbf{0.195} & 	2.65 & 	\textbf{3.30} $\pm$ 0.06  & 	\textbf{0.147} & 	2.62 & 	\textbf{3.52} $\pm$ 0.06 \\
    \bottomrule
	\end{tabular}
	\vspace{-3mm}
\end{table*}

\subsection{Steering vectors update}
To perform DSBF effectively, 
it is necessary to estimate the steering vector  $\hat{\bm{a}}_{i,l,k}$ accurately. 
The proposed method estimates $\hat{\bm{a}}_{1,l,k}$ as the steering vector of the dominant speech source by using principal component analysis (PCA) with an estimated spatial covariance matrix (SCM). The steering vector for the second path $\hat{\bm{a}}_{2,l,k}$ is estimated such that $\hat{\bm{a}}_{2,l,k}$ is orthogonal to  $\hat{\bm{a}}_{1,l,k}$. The $l^2$-norm of each steering vector is set to $1$. The SCM $\bm{R}_{l,k}$ of the dominant speech source is updated in an online manner to track spatial information of speech sources as follows:
\begin{equation}
	\bm{R}_{l,k}=\gamma_{l,k} \bm{R}_{l-1,k} + (1-\gamma_{l,k}) \bm{x}_{l,k}\bm{x}_{l,k}^{H},
\end{equation}
\begin{equation}
	\gamma_{l,k}=1-\mathcal{M}_{l,k}(1-\alpha),
\end{equation}
where $\alpha$ is a forgetting factor. $\alpha$ is set to $0.99$. $\mathcal{M}_{l,k}$ is a time-frequency mask which controls updating of the SCM depending on the 
speech presence at each time-frequency bin.  
To avoid performance degradation in SCM estimation, it is important to accurately estimate
the time-frequency mask, which essentially selects the 
time-frequency bins where speech sources are dominant. Time-frequency mask estimation is done
by reusing the monaural speech enhancement results and additional DNN training and inference is not needed for time-frequency mask estimation. The time-frequency mask is estimated as $
	\mathcal{M}_{l,k}=\min \left( \frac{\lVert \bm{c}_{l,k}  \rVert_{2}}{\lVert \bm{x}_{l,k}  \rVert_{2}} , 1\right), $
where $\bm{c}_{l,k}=[\begin{array}{cc} c_{1,l,k} & c_{2,l,k} \end{array}]^{T}$ is a stereo signal which contains two speech signals. 
When the dominant speech source is absent due to turn-taking, and a new source is present,  the new speech source is initially enhanced in the second path, but since the time-frequency mask $\mathcal{M}_{l,k}$ specifies the time-frequency bins at which the new speech source is active and the new speech source gradually becomes dominant in the SCM. Since the updated steering vectors are used in spatial beamforming with DSBF in a feedback manner,  the new speech source eventually moves to the first path as the dominant source. With such a mechanism, the proposed method track change of the dominant sound source.

\subsection{Perfect reconstruction}
Especially during turn-taking, one speech source can span over outputs in both paths. However, even in that case, under the assumption that the monaural speech enhancement outputs speech sources with no degradation and completely removes noise, the spatial images of speech sources are reconstructed perfectly in the stereo output signal; The final stereo output signal $\bm{c}_{l,k}$ can be written as follows:
\begin{dmath}
\bm{c}_{l,k}  = \left( \sum_{i=1}^{2} \hat{\bm{a}}_{i,l,k} \hat{\bm{a}}_{i,l,k}^{H} \right) \sum_{n=1}^{2} s_{n,l,k}\bm{a}_{n,k} = \bm{P}\bm{P}^{H} \sum_{n=1}^{2} s_{n,l,k}\bm{a}_{n,k}
=\sum_{n=1}^{2} s_{n,l,k}\bm{a}_{n,k},
\end{dmath} 
where $\bm{P}=\begin{array}{cc} [\hat{\bm{a}}_{1,l,k} & \hat{\bm{a}}_{2,l,k} \end{array}]$ and  $\bm{P}$ is a unitary matrix, because $\lVert \hat{\bm{a}}_{i,l,k} \rVert_{2} = 1$ and $\hat{\bm{a}}_{1,l,k} \perp \hat{\bm{a}}_{2,l,k}$, and $\bm{P}\bm{P}^{H}= \bm{I}$ ($\bm{I}$ is the identity matrix).
Thus, in the ideal case, stereo speech sources are perfectly reconstructed with spatial-cue preservation in the final stereo output signal. 
Although the conventional common-gain based method also reconstructs a stereo speech signal perfectly under the condition that the common gain outputs a speech signal without degradation, it does not ensure that two speech sources are enhanced without degradation when there are two speech sources in the input monaural signal. On the other hand, in the proposed method, even when there are two sound sources in the input stereo signal, one speech  signal is emphasized in the input monaural signal in each monaural speech enhancement path. 
Thus, it is possible to focus on enhancement of a relatively small number of speech sources in the monaural speech enhancement and  it can be expected that noise can be suppressed more effectively with relatively less distortion of speech sources in the proposed dual-path structure.

\section{Evaluation}

\subsection{Setup}
We evaluated the proposed stereo speech enhancement framework with objective and subjective  experiments. 
The sampling rate was 16000 Hz. The number of the band gains $N$ was set to $32$. 
We developed our evaluation datasets by using wsj1\_2345\_db \footnote{\url{https://github.com/fakufaku/create_wsj1_2345_db}}, in which  room dimension, source location, microphone location, SNR, and reverberation time are simulated similarly to spatialized wsj0-2mix \cite{deep_clustering2}. 
We added  noise signals which were extracted from CHiME3 dataset  \cite{Chime32015}. 
For speech source signals, 
we used two datasets, WSJ1 \cite{wsj1_dataset} and LibriSpeech ASR corpus \cite{Librispeech2015}. For WSJ1 corpus, fully overlapped mixtures were generated. For the LibriSpeech ASR corpus, the 
overlap of two speech sources was determined by Sparse LibriMix  \footnote{\url{https://github.com/popcornell/SparseLibriMix}}, and the overlap ratio was set to 0.2. 
We call this dataset Stereo Sparse LibriMix. The number of speech sources was set to 2.  We compared methods with a single-path structure and a dual-path structure. In the methods with a  single path structure,  the output signal is obtained as $c_{m,l,k} = z_{1,m,l,k}$.
We also 
evaluated the common-gain method with the non-adaptive steering vectors (common gain w/ NSV) in which the steering vector is fixed to $\hat{\bm{a}}_{1,l,k}=[\begin{array}{cc} 1 & 1 \end{array}]^{T}$ and $\hat{\bm{a}}_{2,l,k}=[\begin{array}{cc} 1 & -1 \end{array}]^{T}$ so as to confirm the
effectiveness of adaptive steering vector estimation in the proposed method. 
 
\subsection{Experimental results}
We use IPD error and ILD error [dB] as objective measures, and mean opinion score (MOS) as a subjective measure. Since DNN based speech enhancement includes nonlinear processing, it is difficult to evaluate ILD and IPD for each source separately. We evaluate the ILD and IPD of the output stereo signal that contains two enhanced speech sources and those of the non-reverberant stereo signal that contains two oracle clean speech sources.
IPD error is defined  as a distance between the IPD of the output stereo signal  
$\phi_{\hat{\bm{c}}}$ and the IPD of the non-reverberant stereo signal  $\phi_{\bm{c}}$ as follows:
\begin{equation}
	\text{IPD error} =\frac{|\phi_{\bm{c}}- \phi_{\hat{\bm{c}}}  |}{\pi},
\end{equation} 
where phase compensation is incorporated. 
The ILD error is defined as the distance between the ILD of the
output stereo signal  $L_{\hat{\bm{c}}}$ and the ILD of the
non-reverberant stereo signal  $L_{\bm{c}}$ as follows:
	\begin{equation}
		\text{ILD error} =|20 \log_{10}L_{\hat{\bm{c}}}-  20 \log_{10}L_{\bm{c}} |.
	\end{equation} 
MOS testing \cite{MOS} was also carried using the crowd sourcing methodology described in P.808 \cite{MOS2,MOS3}. The number of listeners was 10. 
The evaluation results are shown in Table~\ref{tab:eval_result}. 
The proposed common-gain method with the dual-path structure outperformed the discrete-channel processing method and the baseline common-gain method in the single-path structure, and it achieved the best performance in terms of IPD error and MOS for both WSJ1 and Stereo Sparse Librimix datasets.  In the single-path structure, the common gain method with the proposed adaptive steering vector estimation is superior to the baseline common-gain method. 
By comparing the proposed common-gain method in the dual-path structure  with the common gain w/ NSV, the proposed common-gain method achieved better performance than the common gain w/ NSV in terms of IPD error and MOS. The proposed common-gain method is slightly inferior to the common gain w/ NSV in terms of ILD error in WSJ1, but the proposed common-gain method  is superior to the common gain w/ NSV in Stereo Sparse Librimix. ILD error of the proposed common-gain method in the single-path structure is lower than that of the proposed  common-gain method in the dual-path structure, but the proposed common-gain method in the single-path structure is inferior to the proposed common-gain method in the dual-path structure in terms of MOS. It indicates the importance of perfect reconstruction characteristics in the proposed method in the dual-path structure.

\section{Conclusions}
We proposed a stereo speech enhancement technique which combines DNN-based monaural speech enhancement and spatial beamforming.  By using a dual-path structure with a common band gain between channels in each path, the approach preserves the spatial cues of two input speech sources.
The approach also avoids the complex training requirements of dedicated stereo DNN speech enhancement 
models by relying on a pretrained monaural model.  Experimental results show that the method
to be effective under the condition that there are two speech sources.

\bibliographystyle{IEEEbib}
\footnotesize
\bibliography{StereoVoiceFocus}

\end{document}